# Development of Ontological Knowledge Bases by Leveraging Large Language Models

## LE Ngoc Luyen

Université de technologie de Compiègne, CNRS,
Heudiasyc (Heuristics and Diagnosis of Complex Systems),
CS 60319 - 60203 Compiègne Cedex, France
E-mail: *ngoc-luyen.le@hds.utc.fr*

## Marie-Hélène ABEL

Université de technologie de Compiègne, CNRS,
Heudiasyc (Heuristics and Diagnosis of Complex Systems),
CS 60319 - 60203 Compiègne Cedex, France
E-mail: *marie-helene.abel@hds.utc.fr*

## Philippe GOUSPILLOU

Vivocaz
8 B Rue de la Gare, 02200, Mercin-et-Vaux, France
E-mail: *p.gouspillou@vivocaz.fr*

**LE Ngoc Luyen** is a Computer Science researcher and AI engineer, currently serving as an R&D Engineer at Gamaizer.ia and as a research associate at the CNRS UMR HEUDIASYC laboratory, University of Technology of Compiègne. His work focuses on the development of knowledge bases and generative AI services for competency management and learner guidance. He previously contributed to knowledge-graph-based recommender systems for the used-vehicle market, including domain ontologies, user context profile modelling, and semantic similarity methods. He received a PhD in Computer Science from IMT Atlantique for research on discourse representation modelling for French. His research interests include knowledge graphs, ontologies, recommender systems, deep learning, and NLP/LLMs, with applications in mobility, education and skills, crisis management, and finance.

**Marie-Hélène ABEL** is a Professor in the Computer Engineering Department at the University of Technology of Compiègne (France) and has been Head of Department since 2020. From 2003 to 2020, she coordinated the "*Knowledge Engineering and Information Media*" specialization, and from 2008 to 2018, she led the ICI (Information, Knowledge, and Interaction) research team at the CNRS UMR HEUDIASYC laboratory. Her research covers the Semantic Web, Social Web, Knowledge Management (KM), Recommender Systems, and e-learning. She has authored over 250 publications, leads the MEMORAe project (*http://www.hds.utc.fr/memorae*) and participates in numerous research initiatives at regional, national, and European levels. She is also involved in industrial collaborations related to social network modelling, KM and Frugal AI. She has served on scientific committees for conferences in e-learning, knowledge engineering, and KM, and on editorial boards of several journals. She is Associate Editor of the **International Journal on Semantic Web and Information Systems**, and **IET Collaborative Intelligent Manufacturing**. She was President of the IEEE French Section SMC Chapter from 2014 to 2022 and has been an IEEE Senior Member since 2017.

**Philippe GOUSPILLOU** is Head of Innovation at the startup *VIVOCAZ (https://www.vivocaz.fr/)*, which created the first digital vehicle information system in 2017. A graduate of two French engineering schools (ENIB and SUPELEC), Philippe has always been involved in innovation projects in the fields of image processing and networks. He has (co-)founded four pioneering startups in the transmission and analysis of images and associated data in the security and media sectors. From 2013 to 2015, within the company Perfect Memory, he has also collaborated with the Heudiasyc laboratory at the University of Technology of Compiègne on several (European) projects in the field of semantic web technologies and knowledge management applications for audio-visual media related to radio and television broadcasters.

**Abstract:**

Ontological Knowledge Bases (OKBs) play a vital role in structuring domain-specific knowledge and serve as a foundation for effective knowledge management systems. However, their traditional manual development poses significant challenges related to scalability, consistency, and adaptability. Recent advancements in Generative AI, particularly Large Language Models (LLMs), offer promising solutions for automating and enhancing OKB development. This paper introduces a structured, iterative methodology leveraging LLMs to optimize knowledge acquisition, automate ontology artifact generation, and enable continuous refinement cycles. We demonstrate this approach through a detailed case study focused on developing a user context profile ontology within the vehicle sales domain. Key contributions include significantly accelerated ontology construction processes, improved ontological consistency, effective bias mitigation, and enhanced transparency in the ontology engineering process. Our findings highlight the transformative potential of integrating LLMs into ontology development, notably improving scalability, integration capabilities, and overall efficiency in knowledge management systems.

## 1 Introduction

In the current digital age, where information proliferates at an unprecedented rate across diverse domains, the need for effective knowledge management systems has never been more critical. Ontological Knowledge Bases (OKBs) are central enablers of knowledge management systems (KMS), providing the structured, semantic backbone for capturing, organizing, and reusing knowledge across evolving domains (Jurisica, et al., 2004; Zhang, et al., 2011). They support essential KMS objectives such as knowledge formalization, interoperability, and semantic retrieval. As domains become more dynamic, scalable and adaptive knowledge management, supported by OKBs, becomes a critical challenge we aim to address. However, the manual construction and maintenance of ontologies often struggle to keep pace with the dynamic and evolving nature of knowledge acquisition. Ontologies serve as a central mechanism in knowledge management by enabling the formalization, organization, and retrieval of structured knowledge. Integrating Large Language Models (LLMs) into this process enhances scalability and continuity in managing domain knowledge over time, especially in dynamic environments. Traditional approaches typically rely on manual curation by subject matter experts, a process that is both time-consuming and resource-intensive (Ma, et al., 2019; Elnagar, et al., 2022).

Manual ontology development faces several key limitations. First, it demands close collaboration between domain experts and ontology engineers, often resulting in bottlenecks during knowledge formalization (Garijo, et al., 2024). Second, the iterative refinement and validation of ontological structures are laborious, making rapid adaptation to new or shifting domains difficult (Gomes, et al., 2024). Third, manual development is prone to inconsistencies and omissions, especially in large-scale or interdisciplinary ontologies (Huang, et al., 2005). Additionally, tasks such as documentation, test case generation, and aligning the ontology with competency questions require significant human effort and technical expertise. These challenges hinder scalability and reduce the efficiency of ontology engineering. As knowledge domains continue to evolve, the difficulty of capturing and integrating new knowledge into existing OKBs only increases, leading to gaps and inconsistencies (Noy, et al., 2001; Le, et al., 2022).

Recent advancements in Generative AI and particularly the emergence of LLMs offer compelling solutions to these challenges. Trained on massive volumes of textual data, LLMs demonstrate exceptional language understanding and generation capabilities, allowing them to comprehend and produce human-like text across a wide range of domains (Achiam, et al., 2023; Team, et al., 2023; Touvron, et al., 2023). These models excel at capturing nuanced semantic relationships, contextual meanings, and domain-specific knowledge embedded within unstructured texts. By harnessing their power, there is a unique opportunity to automate and enrich ontology development processes - generating dynamic, contextually relevant content that enhances the depth, adaptability, and overall utility of OKBs (Ciatto, et al., 2024) .

With the rapidly evolving landscape of artificial intelligence and knowledge engineering, the integration of OKBs with LLMs stands as a promising frontier for advancing the understanding and utilization of complex information across diverse domains (Thirunavukarasu, et al., 2023; Wu, et al., 2023; Saab, et al., 2024). LLMs are at the forefront of this integration, showcasing their remarkable ability to automate a wide spectrum of tasks, ranging from simplifying natural language interactions such as producing concise summaries and translating text to addressing complex challenges like converting natural language descriptions into code (Vieira da Silva, et al., 2024). The potential of utilizing LLMs to support the development of OKBs, as well as to populate and enrich them, appears promising. Hence, this paper delves into exploring how LLMs can efficiently undertake this task, thereby reducing the need for manual or semi-automated capability modelling efforts.

Specifically, our study focuses on demonstration examples in the vehicle sales domain, presenting a case study that illustrates the synergistic integration of OKBs and LLMs. We demonstrate how leveraging these technologies addresses challenges in ontology development, maintenance, and utilization. Through empirical analysis and practical experimentation, we show how our approach reduces human effort and enhances the accuracy, relevance, and utility of OKBs in the context of vehicle sales. Our contributions include highlighting the potential of LLMs for developing and enriching OKBs, providing a practical case study that demonstrates LLM application in the vehicle sales domain, and discussing associated challenges and opportunities.

By bridging the gap between structured ontologies and unstructured textual data, our findings hold promise not only for enhancing the automotive industry but also for guiding the enrichment of OKBs in various fields. This synergy between structured representation and contextual comprehension afforded by LLMs opens doors to novel approaches in knowledge development and management, decision support systems, and intelligent information retrieval. Thus, this paper lays the groundwork for further exploration at the intersection of AI and knowledge representation, providing valuable insights and guiding future research endeavours in this dynamic domain.

This work contributes to the field of knowledge management by presenting a practical, LLM-assisted methodology for building and maintaining OKBs. To ensure clarity and reproducibility, the paper is organized as follows: Section 2 reviews related literature on OKBs, ontology development, and LLM integration. Section 3 outlines our methodology. Section 4 presents a case study in the vehicle sales domain. Section 5 discusses evaluation results. Finally, Section 6 concludes with implications for knowledge management and future directions.

## 2. Related Work

Similar to numerous domains, the use of LLMs for ontologies, knowledge representation, and their domain-specific applications has surfaced through diverse research endeavours. In this section, we offer an overview of current literature and research efforts regarding Knowledge Management, OKBs and LLMs, emphasizing their distinct roles, capabilities, and the evolving landscape.

### *2.1 Ontologies and Knowledge Management*

Ontologies and the OKBs built upon them play a vital role in KMS. As formal and explicit specifications of shared conceptualizations, ontologies provide a semantic foundation for organizing, contextualizing, and sharing knowledge across organizations (Zhang, et al., 2011). They facilitate semantic consistency across diverse and heterogeneous data sources, support structured knowledge acquisition and exchange, and enable reasoning for more informed decision-making (Jurisica, et al., 2004). Ontologies represent domain knowledge through well-defined classes, relationships, and logical rules, allowing for precise and interoperable descriptions (Kaewboonma & Tuamsuk, 2017). In the context of knowledge management, they help categorize knowledge into four main types: static knowledge, which includes entities and their attributes; dynamic knowledge, which describes processes and changes in state; intentional knowledge, which refers to beliefs, goals, and preferences of agents; and social knowledge, which encompasses roles, norms, and organizational structures. This structured representation supports better alignment between knowledge creation, integration, and practical use (Straccia & Pollo-Cattaneo, 2024).

Ontologies have found wide-ranging applications in KMS across various domains. In healthcare and agriculture, they serve to standardize terminology, thereby supporting clinical decision-making (Elhadj, et al., 2021) and facilitating the sharing of agricultural knowledge (Jonquet, et al., 2018). In the educational sector, ontology-driven systems enhance the organization and retrieval of e-learning resources, contributing to more personalized and effective learning experiences (Abel, 2022). Within the e-commerce domain, ontologies support the development of recommender systems by structuring product information, user preferences, and contextual data, thereby enabling more accurate and personalized recommendations (Le, et al., 2023a; Le, et al., 2023b).

Recent technological advancements have significantly enhanced the development and utility of OKBs within KMS. LLM-based Generative AI facilitates ontology construction through automated concept extraction and enables dynamic updates to knowledge structures (El Idrissi, 2025). In parallel, Big Data technologies benefit from ontologies that provide structure to unstructured data, improving analytics in different domains (Bennett & Baclawski, 2017; Kitchin & McArdle, 2016). Among these advancements, the integration of LLMs stands out by automating labour-intensive tasks such as extracting concepts from unstructured text, identifying semantic relationships, and enriching ontological entities with contextual knowledge (Chang, et al., 2024). This synergy between LLMs and OKBs addresses longstanding challenges in scalability and maintenance while enhancing the adaptability and responsiveness of KMS in evolving environments. To understand how these benefits are realized in practice, we next examine how OKBs are constructed and maintained - focusing on core development methodologies and the role of ontology engineering.

### *2.2 Ontological Knowledge Bases and Development Methodologies*

In the context of our work, OKBs are defined as structured repositories of knowledge, meticulously designed to encapsulate concepts, relationships, and constraints within a specified domain. These repositories are constructed upon ontologies, which provide a machine-readable, formal semantic model. Specifically, ontologies serve as a method of knowledge representation, constituting a collection of concepts within a domain and the intricate interconnections among them (Gruber, 1993). Ontologies[1] rely on explicit formal descriptions, facilitating shared comprehension of information structure, promoting the reuse of domain knowledge, and segregating domain knowledge from operational knowledge.

The utility of ontologies is underpinned by several key reasons: (i) Organizing data: Ontologies provide a structured framework for organizing data by visualizing the inherent concepts and relationships within a domain. By structuring information based on natural structures, ontologies enhance the clarity and accessibility of data, facilitating efficient data management and retrieval; (ii) Enhancing search capabilities: Leveraging ontologies can significantly improve the accuracy and effectiveness of search functions within applications. Unlike traditional keyword-based searches, which may yield limited results, ontological search allows for semantic enrichment, enabling the retrieval of synonyms and related terms that enhance the comprehensiveness of search results; (iii) Facilitating data integration: Ontologies serve as semantic glue, facilitating the seamless integration of disparate data sources and languages. By establishing a common semantic framework, ontologies enable the interoperability of heterogeneous information sources, thereby providing users with a unified entry point to diverse information repositories. This integration enhances the accessibility and usability of data, empowering users to navigate complex information landscapes more effectively (Abiteboul, et al., 2011; Luyen, et al., 2016).

[1] For simplicity, we will use "*ontologies*" and "*ontological knowledge base*" interchangeably.

The development of ontologies encompasses several essential phases, prominently including ontology engineering, which encompasses the creation, refinement, and validation of ontological structures. This iterative process often involves the collaborative efforts of domain experts to meticulously define concepts, properties, and axioms, thereby ensuring accuracy and comprehensiveness (Noy, et al., 2001; Pinto & Martins, 2004). In this context, ontology development methodologies play a critical role by offering a structured and systematic approach to ontology creation, maintenance, and evolution. These methodologies are meticulously designed to uphold the quality, coherence, and practicality of the resultant ontology, while also fostering its consistency and usability. Through the adoption of ontology development methodologies, researchers and practitioners can effectively elevate the overall performance and applicability of their ontologies, thereby enhancing their reliability and value for their intended use cases.

Numerous ontology development methodologies have been suggested in scholarly literature, each characterized by its individual set of guidelines and principles. Among the commonly embraced methodologies are:

- *Methontology* stands as one of the earliest and most extensively applied methodologies in ontology development. It offers a framework that spans the entire ontology development lifecycle, encompassing specification, conceptualization, formalization, integration, implementation, and maintenance stages. A notable feature of *Methontology* is its emphasis on meticulous documentation at every phase of the development process, recognizing its crucial role in ensuring clarity and facilitating future modifications (Fernández-López, et al., 1997). This methodology provides a structured approach that enables ontology developers to systematically navigate through each stage, thereby fostering the creation of robust and sustainable ontological frameworks.

- *Ontology Development 101* (*OD101*) stands as a meticulously crafted methodology tailored to steer ontology engineering endeavours. Emphasizing an iterative approach, *OD101* furnishes step-by-step guidelines and best practices for constructing ontologies, thereby ensuring a methodical and effective pathway to ontology creation (Noy, et al., 2001). This methodology offers a structured framework that enables ontology developers to navigate through the intricacies of ontology development with precision and clarity.

- *The Unified Process for Ontology Building* (*UPON*), proposed by De Nicola, et al. (2005), integrates principles from the Unified Modeling Language (UML) and the Rational Unified Process (RUP). Following an iterative and incremental model, *UPON* encompasses four development phases: inception, elaboration, construction, and transition. This methodology provides a structured framework for ontology creation and refinement, ensuring consistency and systematic development. During the inception phase, stakeholders define the ontology project's scope and objectives. The elaboration phase involves detailed analysis and specification of ontology requirements. In the construction phase, ontology engineers create and implement the ontology using UML-based modelling techniques. The transition phase focuses on testing, deployment, and refinement.

- *The NeON Methodology*, as outlined by Suárez-Figueroa et al. (2011), diverges from conventional approaches by focusing on the creation of ontology networks rather than individual ontologies. This innovative strategy underscores the importance of leveraging and refining existing ontologies, promoting collaboration between ontology engineers and domain experts to enhance the development process. By prioritizing the reuse and re-engineering of ontological resources, *the NeON Methodology* encourages a more comprehensive and efficient approach to ontology development.

- *The Simplified Agile Methodology for Ontology Development* (*SAMOD*), proposed by Peroni (2017), integrates principles from agile software development into the ontology engineering process. It provides a flexible, iterative, and adaptable framework that promotes collaboration among ontology engineers, subject matter experts, and end-users. Utilizing repeated development cycles, *SAMOD* encompasses key stages like planning, design, implementation, and evaluation. This cyclical approach facilitates continuous feedback and refinements throughout the ontology's entire development lifecycle. By embracing agile practices, *SAMOD* enables ontology projects to be more responsive to changing requirements and stakeholder needs, while maintaining a structured process flow. The methodology's emphasis on iteration and feedback loops aims to produce high-quality, well-evaluated ontologies that accurately capture the target domain.

- *Modular Ontology Modelling* (*MoMo*), outlined by Shimizu et al. (2023), is a methodology dedicated to crafting and managing modular ontologies. This approach enables the independent development, maintenance, and reuse of smaller, more manageable ontology modules. *MoMo* promotes collaboration and concentrates on specific sections of the ontology. It entails module identification, design, integration, and evaluation stages, ultimately leading to a streamlined and sustainable ontology development process.

In general, employing an ontology development methodology can significantly enhance ontology development efficiency, promote ontology reuse, simplify ontology maintenance, and foster effective collaboration between ontology engineers and domain experts. Additionally, adopting a standardized ontology development methodology can enhance ontology interoperability and compatibility, facilitating more efficient data integration and knowledge sharing across diverse systems and applications.

The selection of an appropriate ontology development methodology depends on several factors, including the complexity of the ontology, the domain of application, the availability of resources, and the expertise of the ontology engineers. In this study, we focus on the *SAMOD* to construct the framework of our OKBs. We made this decision based on *SAMOD*'s attributes of flexibility, adaptability, and emphasis on collaboration among ontology engineers, domain experts, and end-users. The iterative and incremental nature of *SAMOD*, coupled with its mechanisms for ongoing feedback and refinement, aligns seamlessly with our objective of developing a comprehensive and adaptable ontology capable of evolving over time. In the subsequent section, we will examine and discuss the relevant literature on the progress and developments in the field of artificial intelligence, specifically focusing on the advancements made through the use of LLMs.

### *2.3 The Rise of Large Language Models*

LLMs are advanced deep learning models designed to understand and generate human-like text. They are trained on massive datasets and utilize neural network architectures to learn language patterns, enabling them to perform a wide range of natural language processing tasks. For instance, GPT-3 contains 175 billion parameters, and GPT-4, the latest language model from OpenAI, is significantly larger. These models are trained on enormous datasets comprising text from books, articles, websites, and other sources. Typically based on the Transformer architecture, LLMs employ self-attention mechanisms to capture long-range dependencies in text (Vaswani, et al., 2017). They are then fine-tuned using Reinforcement Learning from Human Feedback, a process that refines their performance through iterative human-guided adjustments (Achiam, et al., 2023). Many LLMs have been developed by leading technology companies, demonstrating their strength in innovative filed of Generative AI. For example, OpenAI's ChatGPT, Google’s Gemini, Meta’s LLama, Microsoft’s Copilot, along with other models like Mistral and Claude, represent significant advancements in this technology. Specifically, LLMs have demonstrated remarkable capabilities, human-level performance on the majority of these professional and academic exams. For instance, models like GPT-3 and GPT-4 have excelled in standardized or advanced exams, including the SATs, GREs, International Mathematical Olympiad, and bar exams (Achiam, et al., 2023; Katz, et al., 2024; Trinh, et al., 2024). LLMs has demonstrated notable success across various specialized domains, including healthcare, legal analysis, education, customer support, software development, and finance (Luyen & Abel, 2025). In healthcare, it aids professionals by comprehending and generating clinical notes while supporting literature reviews and providing accurate responses to medical queries (Thirunavukarasu, et al., 2023; Saab, et al., 2024). In the legal field, LLMs efficiently summarizes legal documents and assists with legal research (Cui, et al., 2023). In education, it provides tutoring support in subjects like mathematics and physics, generates literature reviews, and translates academic texts (Rudolph, et al., 2023; Kasneci, et al., 2023; Le & Abel, 2025). For businesses, LLMs automates customer support, analyses market trends, and generates business reports (George & George, 2023). In software development, it writes and refactors code, identifies bugs, and generates technical documentation (Ahmad, et al., 2023; Mishra, et al., 2024). In finance, LLMs summarizes financial reports, offers portfolio management advice, and helps institutions with regulatory compliance (Wu, et al., 2023; Yang, et al., 2023). Overall, their ability to understand and generate human-like text enables them to perform complex tasks such as reasoning, summarization, translation, problem-solving, and supporting a wide range of professional applications.

Recent research suggests that LLMs have the potential to revolutionize ontology engineering. Their ability to process vast amounts of text data offers new avenues for automating ontology learning, as they can automatically extract knowledge from text sources for ontology population, significantly reducing manual effort (Meyer, et al., 2023;

Trajanoska, et al., 2023; Neuhaus, 2023; Du, et al., 2024; Laurenzi, et al., 2024). Additionally, LLMs can analyse text data to identify new concepts and relationships relevant to a specific domain, leading to more accurate ontologies. They can also enrich existing ontologies with additional information gleaned from textual sources, enhancing the overall richness and usefulness of the knowledge base. For example, the study by Laurenzi et al. (2024) employs LLMs to realize steps for tasks like generating questions, extracting entities, converting data formats, and creating validation rules. They evaluated the effectiveness of their LLM-aided process on real-world data and found that LLMs performed well, particularly in generating relevant competency questions and extracting key terms for the ontology schema. Or the study by Meyer et al. (2023) investigates the potential of LLMs to assist with Knowledge Graph Embedding (KGE) tasks and generate text. The researchers conducted experiments to evaluate whether ChatGPT could aid various KGE activities, including generating SPARQL queries from natural language questions, extracting knowledge from text, and exploring and visualizing existing knowledge graphs. While ChatGPT demonstrated potential in automating some KGE tasks, human oversight and validation remain essential due to the model's tendency to produce errors. The authors Trajanoska et al. (2023) investigate using LLMs like ChatGPT to automatically create knowledge graphs from unstructured text, focusing on sustainability news articles. Comparing the specialized relation extraction model, Relation Extraction by End-to-end Language Generation (REBEL), with ChatGPT, they found that REBEL excelled at identifying entities and relations, while ChatGPT often generated relation phrases instead of concepts. However, with carefully crafted prompts, ChatGPT could generate and populate ontologies, creating a more relevant knowledge graph. Despite limitations like reliance on human assessment and handling large text inputs, the study shows promise for LLMs in knowledge graph embedding tasks, emphasizing the importance of prompt design and post-processing. In general, these studies suggest that LLMs have the potential to revolutionize ontology engineering. However, careful guidance is required to maximize their effectiveness. This includes prompt design, domain-specific fine-tuning, and the integration of post-processing steps to ensure accurate ontology generation. With these considerations, LLMs can enhance tasks like ontology learning, relation extraction, and knowledge graph construction, ultimately leading to more accurate and relevant OKBs.

Ontology development and maintenance have traditionally been laborious and resource-intensive processes, facing several challenges: (i) time consumption: building ontologies with vast amounts of knowledge is a time-consuming process. (ii) expertise dependency: manual development relies heavily on domain expertise and knowledge engineering skills. (iii) maintenance difficulty: keeping ontologies up-to-date with evolving knowledge is a complex task (Ma, et al., 2019; Elnagar, et al., 2022).

OKBs are famous for representing knowledge in a structured manner. They can be applied in many tasks such as question answering, recommendation, and semantic search. However, conventional OKBs are often incomplete, and existing methods often fail to consider textual information. To address these issues, recent research has explored integrating LLMs to augment, populate, or enrich OKBs, enabling them to consider textual information and improve performance in different tasks. In the next section, we delve into our methodology with the support of LLMs in the development, learning, population, and enrichment of ontologies.

## 3. Methodology: LLM-Enhanced Ontology Engineering

In this section, we introduce a methodology that integrates LLMs based on the principles of *SAMOD* (Noy, et al., 2001). The proposed approach leverages the iterative refinement principles of *SAMOD* alongside the powerful knowledge extraction capabilities of LLMs. This hybrid methodology enables efficient and accurate ontology development by combining prompt-based information extraction with agile, incremental refinement. We will detail each step of this methodology, from identifying competency questions to deploying and maintaining the ontology.

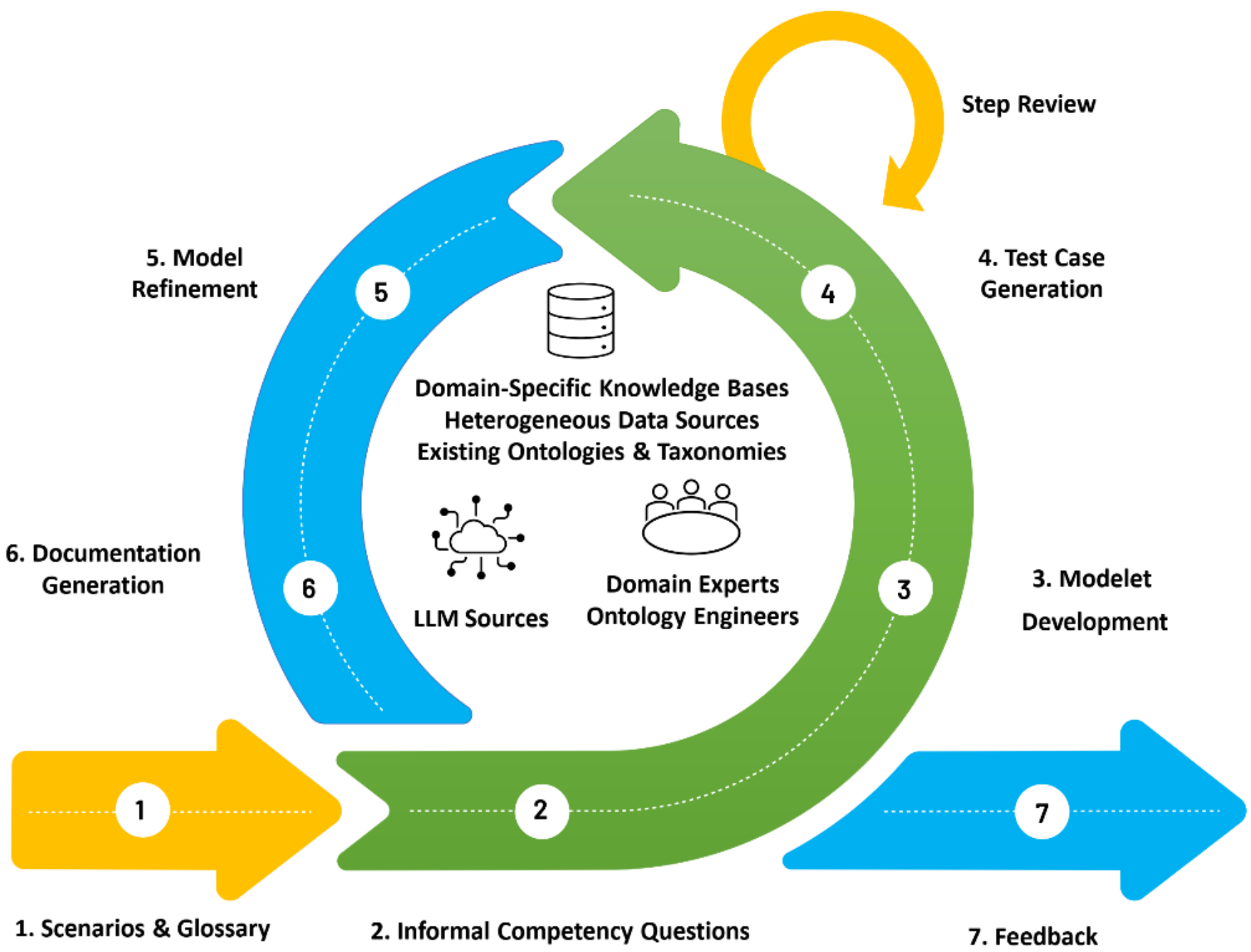

Figure 1: Our proposed methodology

The proposed methodology follows a structured, iterative process illustrated in Figure 1, consisting of seven key steps. It begins with the definition of representative scenarios and the creation of a shared glossary to establish domain understanding. This is followed by the formulation of informal competency questions, which guide the scope and focus of the ontology. The third step involves modelet development, where modular ontology components are created based on identified concepts and relationships. These modelets are then validated through test case generation, ensuring that the ontology can answer the competency questions. The process continues with model refinement, incorporating feedback and ensuring logical consistency. Documentation generation follows, producing human-readable explanations and technical specifications. Finally, a feedback loop ensures continuous improvement by incorporating input from stakeholders. Each step includes a review mechanism to monitor progress and maintain quality. The methodology is collaboratively driven by domain experts and ontology engineers, and is significantly enhanced through the integration of LLMs and resources such as domain-specific knowledge bases, heterogeneous data sources, existing ontologies, and taxonomies.

## *3.1 Development Process*

### *3.1.1 Scenario and Glossary*

The scenario and glossary of terms forms the foundation for any successful ontology development process. Scenarios describe representative situations users might encounter within the target domain, offering concise descriptions and informal, intuitive examples. Rather than technical specifications, these scenarios are relatable stories that illustrate the real-world context and intended use cases for the ontology (van Harmelen, et al., 2002). By considering diverse scenarios jointly crafted by domain experts and ontology engineers, the team gains a clear understanding of the practical applications driving the ontology's requirements. This process helps define the purpose of ontology development and identify the scope for further ontology refinement. Moreover, the glossary of terms serves as the *"language dictionary"* underpinning the ontology. Here, precise definitions for key terms, concepts, and entities relevant to the domain are meticulously documented. Establishing this shared vocabulary ensures all stakeholders - domain experts, ontology engineers, and eventual ontology users - have a consistent, unambiguous interpretation of the core terminology (Fernández-López, et al., 1997). Robust term definitions help prevent misunderstandings and diverging perspectives later in the development lifecycle.

Collaborating with domain experts is crucial for identifying representative scenarios that guide ontology development, and LLMs can provide valuable assistance throughout this process. LLMs can extract common scenarios from analysing domain-specific literature, manuals, and reports, while well-crafted prompts help identify the most relevant use cases - for instance, generating a scenario like *"A user wants to compare the fuel efficiency of different car models from various manufacturers"* for a vehicle model ontology. Moreover, LLMs can refine the scope of each scenario by pinpointing relevant models, parameters, and concise scope statements such as *"The ontology will focus on car brands, models, and their technical specifications for vehicles manufactured between 2010 and 2024"*. They can also document detailed scenario descriptions including actors, actions, and expected outcomes based on initial inputs, generating documentation. This LLM-assisted scenario identification and scoping process keeps ontology development focused on practical, user-centric use cases that address intended users' real needs.

The creation of a glossary that standardizes terminology by clearly defining key concepts, classes, and relationships is essential for ensuring consistent understanding throughout the ontology development process. LLMs can provide valuable support in this endeavour by analysing domain literature, standards, and expert interviews to identify essential terms, generate relevant synonyms and acronyms, and provide accurate definitions with illustrative examples that clarify how concepts interrelate - for instance, describing how *FuelEfficiency* relates to *VehicleModel*. Moreover, LLMs can analyse the evolving glossary, identify potential gaps, and suggest additional terms for inclusion as new concepts and relationships are incorporated. They can also assist in logically organizing and generating structured documentation for the glossary to serve as an accessible reference during subsequent ontology development stages. This LLM-assisted glossary development approach promotes clarity, consistency, and standardization of terminology usage while enabling efficient development by maintaining a living glossary that stakeholders can easily consult.

### *3.1.2 Informal Competency Questions*

The Informal Competency Questions (ICQs) comprise a set of natural language queries designed to transform end-users' information requirements into structured, actionable questions. Serving as a link between user needs and the formal ontology structure, these questions establish a framework for identifying pertinent concepts, relationships, and scope. ICQs play a pivotal role in guiding the development and refinement of the ontology, ensuring its ongoing relevance and alignment with stakeholder needs (van Harmelen, et al., 2002).

The primary purpose of ICQs is to capture essential questions that the ontology should be able to answer. By defining these questions, ICQs help establish the boundaries of the ontology, keeping it focused on relevant topics. They also serve as benchmarks for testing and validating the ontology's structure and its ability to deliver accurate and relevant answers.

Developing ICQs involves several steps, beginning with identifying key stakeholders and understanding the core use cases that the ontology should support. For instance, in a vehicle model ontology, stakeholders might include car enthusiasts, automotive journalists, and dealers, while critical use cases could range from comparing car models to understanding fuel efficiency and safety features. Once stakeholders and use cases are identified, the next step involves formulating initial ICQs based on user needs and domain knowledge. These questions should be written in natural language to ensure comprehensibility by non-technical stakeholders. Examples might include, *"What are the specifications of the 2020 Toyota Corolla?"* or *"Which car models have the best fuel efficiency in 2022?"*.

After generating initial questions, the ICQs are iteratively refined and expanded in collaboration with domain experts. This process ensures coverage and addresses variations in terminology. For example, *"Which car models are the most fuel-efficient?"* could be expanded to include synonyms like *"fuel economy"* or related questions like *"Which cars offer both fuel efficiency and high horsepower?"*.

LLMs can significantly enhance the development of ICQs by automating the generation, refinement, and expansion of questions. LLMs can analyse domain-specific literature, user queries, and manuals to generate initial ICQs, such as *"Generate a list of questions that car enthusiasts might ask about fuel efficiency"*. They can also suggest synonyms, related terms, and variations of questions for coverage. For instance, given *"What are the best electric cars?"* an LLM might suggest variations like *"Which electric cars have the longest range?"* or *"What are the most affordable electric cars?"*. Additionally, LLMs can help categorize and prioritize ICQs into logical groups based on frequency and importance, like grouping technical specification questions separately from safety features. Finally, LLMs can iteratively refine and expand ICQs based on new domain information by incorporating feedback and analysing recent literature, suggesting new questions that reflect emerging trends and terminology.

### *3.1.3 Modelet Development*

During modelet development, the initial ontology structure is derived from collaborative workshops with domain experts, analysis of existing domain ontologies, and domain-specific resources such as academic literature, technical reports, and structured databases. LLMs significantly enhance this step by automatically extracting relevant concepts and proposing initial ontology structures through targeted prompts. For example, LLMs analyse automotive domain literature to suggest core classes (e.g., *VehicleType*, *FuelEfficiency*) and key relationships (e.g., *hasBrand*, *hasSpecification*), rapidly generating initial modelets for validation and refinement.

A modelet, in the context of ontology engineering, is a small, self-contained module or building block that encapsulates a specific aspect or subset of domain knowledge within a larger ontology. It functions as a modular component that can be independently developed, tested, and refined prior to integration into the primary ontology model. Therefore, modelets facilitate a divide-and-conquer strategy in ontology development, enabling the ontology to be built incrementally by assembling these smaller, manageable knowledge representation units.

In our proposed methodology, the modelet development step involves creating a preliminary, lightweight version of the ontology, referred to as a modelet. This modelet serves as a prototype to help visualize, test, and refine the ontology's structure before full-scale development. It allows ontology engineers to identify key concepts, relationships, and hierarchies early in the development process. The primary purpose of the modelet development step is to build a skeleton ontology that covers the core competency questions while remaining flexible for further refinement. This preliminary structure includes basic classes, properties, and relationships, providing a framework to test and validate the ontology's alignment with stakeholder needs.

Developing a modelet typically involves several steps, beginning with defining core classes and relationships based on the scenarios and ICQs outlined earlier. For instance, in a vehicle model ontology, initial classes might include *VehicleBrand*, *VehicleModel*, and *FuelEfficiency*, with relationships like *hasBrand*, *hasModel*, and *hasSpecification*. These initial elements form the foundation of the modelet. Next, the modelet is populated and enriched with sample instances to validate the structure. For instance, instances of *VehicleModel* could include *"Toyota Corolla"* and *"Honda Civic"*, while the *VehicleBrand* class would include *"Toyota"* and *"Honda"*. Creating these instances helps verify that the modelet can adequately represent real-world data.

In this step, LLMs can be employed to analyse domain-specific literature, user queries, and manuals to identify core classes and relationships for the modelet. For example, prompting an LLM with *"List key concepts and relationships for vehicle models"* can produce a preliminary list of relevant classes and relationships. LLMs also assist in creating concise and accurate definitions for each class and relationship. Given the prompt "*Define the class VehicleBrand*", the model can produce a definition like *"VehicleBrand: A company that manufactures or distributes vehicles, such as Toyota or Ford"*.

Additionally, LLMs iteratively refine the modelet by analysing competency questions and suggesting new classes, relationships, or attributes. For instance, based on the prompt *"Suggest additional properties for the class VehicleModel"*, the LLM might generate properties like *hasSafetyFeatures* or *hasEngineType*. In summary, leveraging LLMs enhances this process by automating the identification of core concepts, relationships, and instances while providing accurate definitions and testing queries.

#### *3.1.4 Test Case Generation*

In our methodology, test cases are systematically derived from the set of ICQs. Each ICQ is translated into a corresponding SPARQL query used to validate the ontology's ability to return expected results. LLMs are instrumental in automating this translation process: they parse the intent behind each ICQ, identify relevant ontology classes and properties, and generate the appropriate SPARQL structure. For instance, an ICQ such as *"Which car models are most fuel-efficient?"* is converted by the LLM into a query targeting the *VehicleModel* class and its *hasFuelEfficiency* property. LLMs also assist in generating edge-case test queries and alternative formulations to ensure robustness. These SPARQL test cases are run iteratively throughout development to detect coverage gaps, inconsistencies, or logical errors in the ontology design.

The test case generation aims to ensure that the ontology meets the defined competency questions and delivers accurate results. Test cases serve as practical examples or queries that validate the ontology's structure and its ability to provide relevant information (Vrandečić & Gangemi, 2006). By systematically testing the ontology against these cases, ontology engineers can identify gaps, refine the ontology, and ensure alignment with stakeholder needs. The significance of test cases lies in their ability to serve as a step review (i.e., a checkpoint) in the development of modelets, allowing for ongoing integration into the final model or reverting to earlier stages to incorporate missing concepts or relationships.

Test cases are crafted from ICQs, encompassing a spectrum of use cases and user needs to assess the ontology's capability to address pertinent inquiries, establish expected outcomes as a validation benchmark, and ensure alignment of ontology elements with real-world requirements. Throughout each stage of modelet development, test cases are scrutinized against ICQs using sample queries to pinpoint any gaps or areas necessitating refinement. For instance, a competency question such as or *"Which car models boast the highest fuel efficiency?"* would be converted into a SPARQL query to gauge the model's accuracy in delivering responses.

The support of LLMs for test cases is expressed through translating ICQs into formal SPARQL queries and identifying edge cases that challenge the ontology's ability to handle unusual scenarios, such as or *"Vehicle Models with unknown fuel efficiency"* or *"Vehicle Models with exceptionally high efficiency"*. LLMs can suggest sample instances for classes based on domain knowledge, such as generating *"Peugeot"* and *"Tesla"* for the *Vehicle* class. LLMs can help translate ICQs into SPARQL queries to test the modelet. For instance, given the question *"Which car models have the best fuel efficiency?"* an LLM can generate a SPARQL, as shown in Query Box 1.

```
SELECT ?carModel ?fuelEfficiency
WHERE {
 ?carModel ex:hasFuelEfficiency ?fuelEfficiency .
 FILTER(?fuelEfficiency > 30)
}
```

Query Box 1: A SPARQL query generation for the question “*Which car models have the best fuel efficiency?*”

Moreover, LLMs can suggest variations of test cases for coverage, as *“Find cars with automatic emergency braking”* or *“Identify cars with lane-keeping assist”*. They can iteratively refine test cases based on ontology changes and new competency questions, continuously adapting queries and expected results to reflect the evolving ontology.

### *3.1.5 Model Refinement*

The model refinement process is inherently iterative, involving multiple cycles of analysis, update, and validation. Each iteration begins with a review of the ontology’s ability to answer the defined competency questions. If gaps are identified - such as missing classes, insufficient relationships, or misaligned domains - these are addressed by extending the ontology’s structure. Logical consistency is also checked, ensuring subclass hierarchies are sound, property restrictions are valid, and naming conventions are coherent. LLMs assist in suggesting missing elements based on literature analysis and in identifying semantic inconsistencies through prompt-based review of the evolving ontology. Domain experts participate in this process by validating proposed additions and resolving ambiguities. This loop continues until the ontology satisfies both formal correctness and practical relevance.

In the model refinement stage, our focus shifts to enhancing and polishing the preliminary ontology model developed from the modelets. This process involves integrating and improving the structure, relationships, and completeness of the ontology to ensure it accurately represents the domain and effectively addresses competency questions. One of the primary objectives of model refinement is to enhance class and relationship definitions. This entails refining the definitions of existing classes and relationships to ensure clarity and consistency, while also ensuring accuracy and alignment with the domain’s terminology. Additionally, identifying and filling gaps is essential for ontology development. This means finding missing concepts, relationships, or properties that need to be included and expanding the ontology by adding new classes, properties, and relationships to cover unaddressed aspects of the domain.

Logical consistency is another key goal of model refinement. Verifying and refining hierarchical structures helps maintain accurate subclass relationships, while ensuring that domain and range restrictions are correctly applied through *ABox* data and SPARQL. Alignment with competency questions is also critical. The ontology needs to be refined so it can accurately answer all competency questions, requiring adjustments to classes, properties, and relationships to align with user needs.

In general, this stage is highly collaborative, drawing upon the expertise of domain experts, ontology engineers, and the capabilities of LLMs. Domain experts validate the proposed changes and ensure they accurately represent the domain. Ontology engineers propose modifications to the model’s structure, such as adding new entities, refining relationships, or adjusting axioms to address identified shortcomings. LLMs can support collaborative tasks such as identifying gaps, expanding definitions, and improving logical consistency within the ontology. LLMs aid in refining class and relationship definitions by analysing the existing ontology and suggesting more precise definitions. They excel in identifying and filling gaps within the ontology by analysing domain-specific literature and manuals. By prompting an LLM to list additional properties for a class like *Vehicle*, new attributes such as *hasSafetyFeatures* or *hasEngineType* can be generated, enhancing the ontology's completeness. Additionally, LLMs help to improve logical consistency by suggesting refinements to hierarchical structures and domain-range restrictions. For instance, when presented with a subclass relationship like *ElectricVehicle* under *Vehicle*, an LLM could help refine the hierarchy to ensure logical consistency. Lastly, LLMs align the ontology with competency questions by analysing informal queries and suggesting changes to ensure accurate answers. If prompted with a question like *"Which cars have automatic emergency braking?"* an LLMs recommend adding a *hasSafetyFeature* property to the *Vehicle* class.

### *3.1.6 Document Generation*

To support documentation generation, LLMs are used to automatically produce natural language descriptions for ontology classes, properties, and relationships. For each entity, LLMs can generate definitions, usage examples, and human-readable annotations (e.g., *rdfs:comment* and *rdfs:label*) based on the conceptual model and existing ontology structure. These descriptions are directly embedded into the ontology, improving its clarity and maintainability, especially when rendered in ontology editors such as Protégé. Additionally, LLMs assist in drafting summary reports and ontology usage guides by interpreting schema components and structuring the content into coherent technical narratives. This greatly reduces the manual documentation workload and promotes semantic consistency across all elements.

In this stage, our focus shifts towards crafting documentation that describes the ontology's structure, components, and practical application. This documentation serves as a reference guide catering to various stakeholders, such as domain experts, ontology engineers, and end-users. Our primary goals here are to ensure clarity, accessibility, and usability of the ontology by furnishing detailed documentation. Ontology engineers collaborate closely with domain experts to ensure the documentation accurately mirrors the ontology's structure and semantics. Furthermore, they engage with end-users to gather feedback and integrate any necessary revisions into the documentation. Clear and comprehensive documentation functions as a guiding tool for users, empowering them to grasp the model's structure, objectives, and crucial elements. This comprehension leads to enhanced understanding, decreased maintenance costs, and enriched collaboration between developers and users. Effective documentation entails the inclusion of formal specifications featuring detailed technical descriptions of the model using the Web Ontology Language (OWL), alongside annotated entities and relationships with labels (e.g., *rdfs:label*, *rdfs:comment*), facilitating the generation of human-readable documentation from the ontology source by various tools (Peroni, et al., 2012).

LLMs can play a significant role in automating the document generation process. LLMs excel in natural language generation tasks, making them well-suited for creating detailed and coherent documentation. They can analyse the ontology's structure, relationships, and definitions to generate descriptive text that explains each component. For example, given a class like *Vehicle*, an LLM can automatically generate a detailed description that includes its attributes, relationships, and relevant examples. Furthermore, LLMs can assist in organizing the documentation by generating section headings, tables of contents, and indexes. They can also help in formatting the document to ensure consistency and readability. By leveraging LLMs for document generation, ontology engineers can efficiently manage the documentation process, saving time and effort while ensuring the quality and comprehensiveness of the final product.

### *3.1.7 Feedback*

To systematically incorporate stakeholder feedback, our methodology integrates structured mechanisms such as expert interviews, annotated usage logs, and open-ended feedback forms. Free-text inputs from domain experts, ontology users, and developers are processed using LLMs to extract actionable insights. Specifically, LLMs are prompted to identify recurring themes, prioritize pain points, and summarize suggestions. This enables efficient analysis of large volumes of qualitative feedback. The extracted information is then used to revise competency questions, refine class structures, or add missing properties. This feedback loop is conducted iteratively, ensuring that ontology evolution remains closely aligned with stakeholder expectations and domain-specific requirements.

The feedback stage of ontology development helps to refine and improve the ontology based on real-world usage and stakeholder input. It serves as a mechanism for continuous improvement and adaptation to evolving needs and requirements. Stakeholders, including domain experts, ontology engineers, and end-users, provide feedback on various aspects of the ontology, such as its structure, content, usability, and effectiveness in addressing their needs. The feedback can be solicited through surveys, interviews, user testing sessions, or direct communication channels, depending on the context and preferences of the stakeholders involved (Norris, et al., 2021).

Domain experts contribute their expertise to evaluate the ontology’s alignment with domain-specific knowledge and requirements. They provide insights into whether the ontology accurately represents the domain and captures relevant concepts, relationships, and constraints. Ontology engineers leverage their technical knowledge to assess the ontology’s adherence to best practices, standards, and design principles. They also evaluate the ontology’s scalability, interoperability, and maintainability, identifying areas for improvement and optimization. Additionally, end-users provide feedback on the ontology’s usability, accessibility, and effectiveness in addressing their needs and objectives. Their input helps identify usability issues, user experience challenges, and areas where the ontology may fall short in meeting their expectations. End-user feedback can uncover usability barriers, ambiguities in terminology, or gaps in functionality that need to be addressed to enhance the ontology’s utility and user satisfaction.

LLMs can help to process large volumes of textual feedback, extract key themes, sentiments, and suggestions, and identify recurring patterns or issues raised by stakeholders. By analysing feedback from diverse sources, LLMs can help prioritize improvement efforts, identify areas of consensus or disagreement among stakeholders, and generate recommendations for ontology refinement.

### *3.2 Leveraging Existing Resources and LLMs Integration*

The development of any ontology necessitates the effective utilization of a diverse range of existing resources, including domain-specific knowledge bases, heterogeneous data sources (both structured and unstructured), and relevant existing ontologies or taxonomies from the target domain. Tapping into these valuable resources not only accelerates the ontology creation process but also ensures the resulting model accurately captures and aligns with established domain knowledge and standards.

Table 1: Prompting Technique Comparison

| **Technique** | **Descriptions** | **Examples** |
|---|---|---|
| Zero-shot Prompting | Instructs LLMs with a natural language prompt without any specific training examples. | Generate a list of key concepts and relationships for modelling vehicles. Or, describe the features of an ideal family vehicle. |
| Few-shot Prompting | Provides a few examples to guide LLMs' response. | Given examples like *"Vehicle hasEngine Engine"* and *"Car isA Vehicle"*, suggest other properties and hierarchies for a vehicle model. |
| Chain-of-Thought Prompting | Breaks down the reasoning process step-by-step for LLMs to follow. | To model a user's vehicle preferences: 1. Identify core attributes like vehicle type, brand, budget. 2. Consider contextual factors like commute, climate, family size. 3. Organize into a structure with classes, properties, relationships. |
| Self-Consistency Prompting | Encourages LLMs to generate responses that are consistent with itself across prompts. | List vehicle features a user may want. Now suggest a structure to represent those user preferences in a consistent way across different user profiles. |
| General Knowledge Prompting | Injects general factual knowledge into the prompt to improve grounding. | Vehicle brands include *Toyota*, *Honda*, *Ford*, etc. Vehicle types include sedan, SUV, truck, etc. With this background, propose a concept model for capturing user vehicle preferences. |
| Prompt Chaining | Uses multiple sequential prompts to guide LLMs towards a more complex or nuanced response. | 1. List core concepts for a vehicle ontology 2. Now suggest properties for the Vehicle class 3. Finally, outline a hierarchy from Vehicle to specific models |
| Retrieval Augmented Generation | Combines retrieval of relevant information from external sources with generation of text. | Retrieve excerpts from vehicle manuals and user forums, then generate concepts, properties and definitions based on this context. |

By integrating domain-specific knowledge bases, ontology engineers can incorporate expert-curated information, well-defined concepts, and validated relationships directly into the ontological model. Moreover, the ability to seamlessly ingest and process heterogeneous data sources, ranging from structured databases to unstructured text documents, technical manuals, or research publications, is crucial for knowledge acquisition. This facilitates the extraction of relevant concepts, entities, and relationships that may be scattered across diverse information repositories, ensuring a holistic and well-rounded ontological representation. Additionally, leveraging existing ontologies and taxonomies within the target domain provides a solid starting point and a frame of reference for the development process. These pre-existing models can be extended, refined, or mapped to the new ontology, fostering interoperability and knowledge reuse across different applications and systems

Establishing a collaborative platform enables domain experts, ontology engineers, and end-users to contribute their expertise and feedback seamlessly. Effective communication is facilitated through the use of shared terminology and visual representations of the ontology, enhancing clarity and mutual understanding among stakeholders. Additionally, implementing version control mechanisms enables the tracking of changes and maintains transparency throughout the ontology development process. By adhering to a well-established development process like the one proposed in this methodology, collaboration standardizes workflow procedures and ensuring that the ontological model accurately reflects domain-specific nuances, complies with industry standards, and fulfils the specific requirements of its intended applications and user base.

Leveraging LLMs allows for time savings and enhances the process of mining existing resources for ontology development based on different prompting techniques, as listed in Table *1*. With their remarkable natural language understanding capabilities, LLMs function as knowledge bases, aggregating information from diverse sources across the internet. They serve as powerful tools capable of extracting relevant concepts, entities, and relationships from various unstructured data sources like research papers, technical documents, and domain-specific corpora. By analysing these textual resources, LLMs identify and propose ontological structures, axioms, and relationships that accurately represent the knowledge contained within the source material. Moreover, LLMs automate tasks related to knowledge extraction, ontology population, and terminology standardization. Their language processing capabilities facilitate the efficient extraction of key terms, definitions, and semantic associations from existing knowledge bases, taxonomies, and domain-specific resources, seamlessly integrating them into the evolving ontological model.

The integration of LLMs into the ontology development workflow enables ontology engineers to harness the capabilities of these advanced language models, automating and streamlining tasks while extracting knowledge from diverse sources and generating supporting artifacts. However, this process must be complemented by close collaboration with domain experts, who provide validation, refinement, and real-world grounding to maintain the ontology’s integrity and alignment with domain-specific requirements. In the next section, we explore a specific use case of ontology development using the principles outlined in the proposed methodology.

## 4. Case Study: User Context Profile Ontology

We initiate a case study with a scenario centred around user profile management within the domain of vehicles (Le, et al., 2023a). In today's digital landscape, tailoring experiences for users as they interact with diverse systems and applications is paramount. To achieve this, individual user profiles are intertwined with corresponding contextual information, culminating in the formation of a user contextual profile (Li, et al., 2019). A user profile comprises a compilation of personal details and preferences pertaining to a specific user, encompassing demographic data, interests, past behaviours, and other attributes that delineate the user's identity within a system or application (Maria, et al., 2007). For instance, consider *Henri*, who seeks to purchase one vehicle for professional use and another for familial purposes, given his family with two children. Consequently, *Henri* may possess two distinct profiles, each tailored to different needs and preferences. Contextual information associated with each profile pertains to the situational factors surrounding a user's interaction with a system or application (Tamine-Lechani, et al., 2010). This encompasses factors such as location, time, device utilized, and even the user's current activity or emotional state. In our example, each of *Henri's* two profiles entails unique contextual information reflective of his preferences, requirements, and interactions within the application. As a result, a user contextual profile amalgamates the user profile with the contextual information specific to that profile, thereby fostering an understanding of the user's needs and preferences across various scenarios. For instance, during the quest for a professional vehicle, the user contextual profile may prioritize attributes like fuel efficiency, compact size, and reliability, while in the pursuit of a family vehicle, it may emphasize safety features, ample seating, and cargo space.

In this context, establishing a User Context Profile Ontology (UCPO) helps to achieve effective user management that can be reused and applied across diverse domains such as e-learning (Kritikou, et al., 2008), e-commerce (He & Fang, 2008), recommender systems (Le, et al., 2022; Le, et al., 2023c), personalized web services (Sieg, et al., 2007), context-aware applications (Rimitha, et al., 2019), and among others (Muñoz & Cardinale, 2021; Kourtiche, et al., 2020; Le, et al., 2023d; Le, et al., 2023e). Utilizing agile software development principles and harnessing the capabilities of LLMs, the proposed methodology facilitates the efficient creation of a comprehensive and flexible UCPO. This enables the UCPO to evolve dynamically, accommodating ongoing updates and modifications driven by changes in user profiles and contextual information. Such adaptability is achieved by adopting an agile approach to ontology development, where the ontology undergoes incremental construction and refinement, continuously informed by user feedback. We outline our development process, typically consisting of seven stages condensed into sections including preliminaries, development of the UCPO, and experiments and evaluations, providing a concise and comprehensible framework for UCPO design and enhancement. In the subsequent section, we delve into the use case scenario and design foundations, defining the objectives, scope, and ICQs for the UCPO.

### *4.1 Use Case Scenario and Design Foundations*

In light of the scenario, the primary goal of UCPO development is to create a standardized depiction of user information and contextual data for individual user profiles. This ontological user modelling is pivotal for fostering personalized systems and services, such as tailored information retrieval, adaptive user interfaces, personalized recommendations, and more. By establishing a standardized framework for user data, context, and profiles, UCPO enhances interoperability and empowers it to adjust to evolving user needs and preferences seamlessly. The scope of the UCPO is broad, covering various aspects of users and their surroundings, like demographics, interests, and the context of their devices and networks, physical and social environment, and time. It's designed to be flexible and can adapt to new or changing user information and environmental factors.

Table 2: Examples of the Glossary of Terms

| Term | Interpretation |
|---|---|
| User Profile | A compilation of personal details and preferences pertaining to a specific user within a system or application, encompassing demographic data, interests, past behaviours, and other attributes that delineate the user's identity. |
| User Context | Situational factors surrounding a user's interaction with a system or application, including location, time, device utilized, and the user's current activity or emotional state. |
| User Profile Context | The integration of a user profile with contextual information specific to that profile, fostering a comprehensive understanding of the user's needs and preferences across various scenarios. |

In developing the UCPO, we consult a variety of information sources and references such as academic articles, conference proceedings, textbooks, and online resources related to ontology engineering, user modelling, and personalization. We also leverage the capabilities of LLMs to extract relevant concepts and terms, as illustrated in Table *2*. Moreover, we utilize an LLM to analyse a corpus of research papers focusing on user context and personalization within the automotive industry. By deploying the LLM, we can automatically identify key concepts such as user preferences, vehicle usage patterns, and environmental factors influencing user interaction with vehicles. This automated analysis significantly expedites the identification of pertinent information and minimizes the manual effort required for literature review. Additionally, consulting domain experts, ontology engineers, and ontology users provides valuable insights into the specific user characteristics and contextual factors that should be incorporated into the ontology. By combining both manual review and automated analysis with LLMs, we ensure a comprehensive and accurate representation of user context in the UCPO.

Through collaboration among various stakeholders and leveraging the scenario alongside LLM support, we conduct analysis and propose a set of natural language questions. As depicted in Table *3*, we present ICQs aimed at defining the key characteristics of the user contextual profile within the vehicle sales domain. These questions inform the development of the UCPO.

With the establishment of a clear definition of the goal, scope, glossary of terms, and information sources, we can align the objectives of the UCPO development among domain experts and ontology engineers. This shared understanding lays the foundation for the subsequent phases of the development process, where the essential elements of the UCPO are meticulously crafted and implemented to achieve a robust and flexible version. In the upcoming section, we will delve into these main phases, outlining the different stages of the UCPO development.

Table 3: Examples of informal competency questions

| ID | Questions |
|---|---|
| CQ01 | What is demographic information of the user? |
| CQ02 | What is the user’s preferred vehicle type? |
| CQ03 | What is the user’s budget for a vehicle purchase? |
| CQ04 | Which particular vehicle models are favoured by the user? |
| CQ05 | What is the user’s typical commute distance? |
| CQ06 | What is the user’s preferred vehicle brand? |
| CQ07 | What are the primary use cases for a particular vehicle model? |
| CQ08 | What is the user’s preferred vehicle transmission type? |
| CQ09 | What safety features are important to the user? |
| CQ10 | What is the user’s preferred fuel type (gasoline, electric, hybrid, etc.)? |
| CQ11 | Does the user require cargo/towing capacity for their vehicle? |
| CQ12 | What infotainment and technology features are desired by the user? |
| CQ13 | What climate/weather conditions does the user’s driving environment have? |
| CQ14 | How many passengers does the user need to accommodate in their vehicle? |
| CQ15 | Does the user want new or used/certified pre-owned vehicles? |

### *4.2 Ontology Construction with Modelets*

Our methodology follows a structured seven-stage process inspired by agile and test-driven ontology development approaches. This includes scenario analysis, glossary creation, competency question formulation, modelet construction, test case generation, model refinement, and documentation. LLMs support each step to enhance scalability and semantic completeness. These stages are iteratively conducted, ensuring continuous validation and feedback incorporation.

Considered the core of the ontology engineering process, the development of UCPO involves several iterations of modelet development, test cases, and model refinement until a comprehensive and adaptable ontology is achieved, as outlined in our proposed methodology. Our methodology promotes an iterative process that aims to construct the final model through incremental steps. A key component of this process is the development of modelets, which are standalone models describing specific aspects of the domain. These modelets facilitate initial conceptualization without being constrained by the model established in the previous iteration of the process.

We have structured the development of the UCPO into two distinct modelets. The first modelet concentrates on capturing static and permanent user characteristics. In contrast, the second modelet integrates both static and dynamic information, addressing broader aspects of the user profile and the contextual factors that influence it.

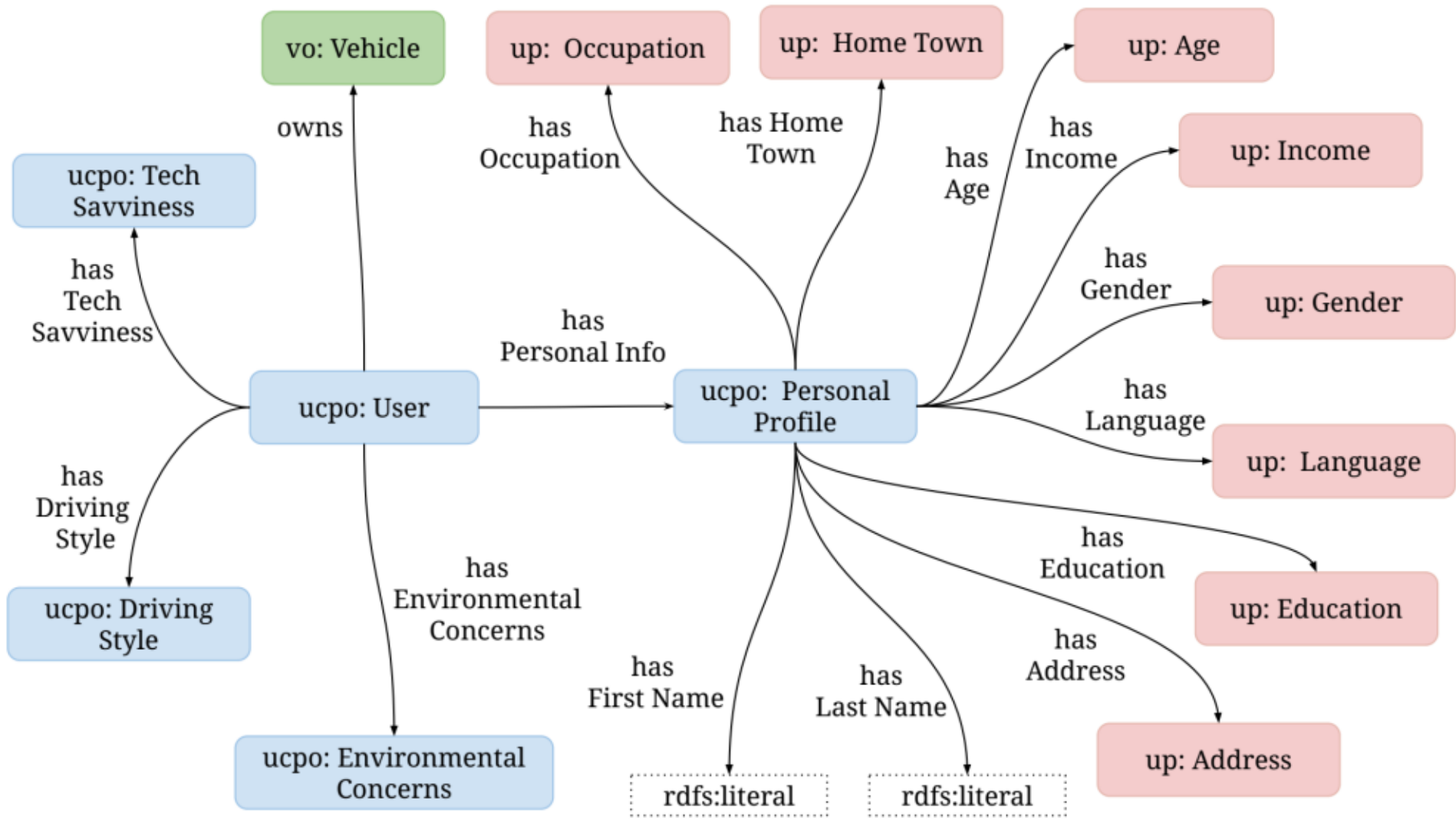

Figure 2: Modelet 1: Ontological structure for modelling static and demographic user information. This modelet integrates three ontology modules: (i) the User Profile Ontology (red, prefix *up*), capturing core demographic attributes (e.g., gender, age, occupation, income), (ii) the User Contextual Profile Ontology (blue, prefix *ucpo*), supporting contextual modelling of user profiles, and (iii) the vehicle domain ontology (green, prefix *vo*), linking user profiles to vehicle-related concepts.

The first modelet, illustrated in Figure 2, is dedicated to capturing stable and permanent user characteristics - such as demographic and social information - that are central to constructing a reliable user profile. These characteristics include core attributes such as *Gender*, *Age*, *Address*, *Occupation*, *Income*, *Language*, *NumberOfChildren*, *MaritalStatus*, and *Education*. Specifically, *Gender* refers to a person's biological identity as male or female; *Age* quantifies the user's age, typically in years; *Address* pinpoints the user's physical location, offering contextual insights into their environment; *Occupation* relates to their professional engagement, shedding light on daily routines and transportation needs; *Income* reflects financial status, influencing purchasing power and preferences; *Language* denotes the preferred communication medium, affecting content types and recommendations; *NumberOfChildren* indicates the number of children, impacting vehicle preferences and requirements; *MaritalStatus* reveals marital status, offering insights into family structure and needs; and *Education* represents the highest academic achievement, influencing interests and preferences. In previous research, user information has been systematically organized into distinct classes, providing a structured approach to categorizing this data. Building on the foundational work cited in (Maria, et al., 2015), our study delineates these essential classes of user information. We have introduced a class named *PersonalProfile*, which serves as a superclass encompassing all related demographic and social information classes. This class also includes attributes such as the user's first and last name.

Additionally, we are exploring new classes that reflect user attributes related to *DrivingStyle*, *TechSavviness*, and *EnvironmentalConcerns*, suggested by leveraging LLMs. For *DrivingStyle*, we capture data on daily mileage, typical driving conditions (e.g., city, highways, off-road), and the primary use of the vehicle (e.g., commuting, leisure, business), enabling tailored vehicle recommendations based on durability, fuel consumption, and functionality to suit the customer's lifestyle. The *TechSavviness* class measures the user's comfort level and interest in technology, particularly with automotive innovations such as autonomous driving features, in-car entertainment systems, and advanced vehicle safety systems, facilitating the recommendation of vehicles that align with the user's technological preferences and enhancing satisfaction and usability. *EnvironmentalConcerns* assess the importance the user places on environmental impact, including their interest in electric vehicles or hybrid technologies, crucial for targeting customers with eco-friendly vehicle options and aligning with trends in environmental sustainability. These new classes aim to provide a more nuanced understanding of users, allowing for more precise and personalized vehicle recommendations.

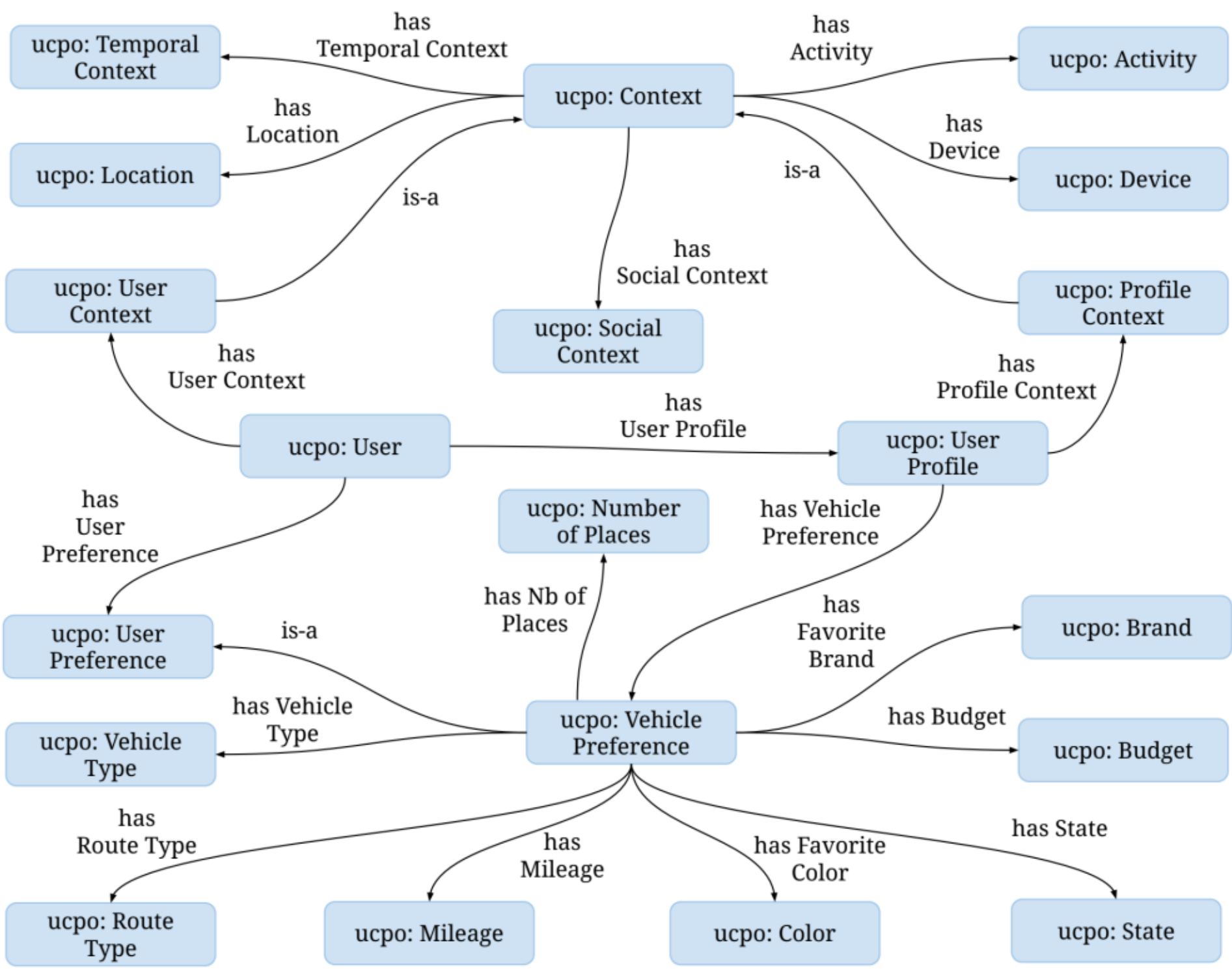

Figure 3: Modelet 2: Ontological structure for modelling user preferences and contextual information. This modelet centres on the relationship between the user, their profiles, and contextual elements - grouped into *TemporalContext*, *Location*, *Activity*, *Device*, and *SocialContext*. It also includes classes representing vehicle-related preferences (e.g.,

*VehicleType*, *Budget*, *Brand*), enabling fine-grained, context-aware personalization of recommender systems.

The second modelet delves into the dynamics of how a user interacts with systems or applications, mapping out the structure and relationships between the user, their profile, and their broader context. This context comprises a myriad of elements such as location, time of day, day of the week, specific date and time, seasonal changes, device specifics (e.g., type and operating system), user activities both past and current, as well as the social environment (e.g., social media interactions and the influence of social circles). Illustrated in Figure 3, our approach categorizes these elements into detailed classes like *TemporalContext*, *SocialContext*, *Location*, *Activity*, and *Device*. We have further refined our model to include *UserContext* and *ProfileContext* subclasses that capture general and specific contextual data, respectively. Additionally, we enable users to specify their preferences and customize their profiles through the *Preference* and *UserProfile* classes, enhancing their engagement by aligning system interactions with their goals and circumstances.

In order to address the needs and preferences of users in the vehicle sales domain, we have designed a set of preferences that relate to the user's desired vehicle. These classes include *VehicleType*, which specifies the type of vehicle, such as a sedan, SUV, or truck; *RouteType*, which details the user's preferred driving routes, whether highways or city streets; *Mileage*, which captures the desired mileage range; *Color*, which denotes the user's favourite vehicle color; *NumberOfPlaces*, which determines the seating capacity needed; *State*, which identifies where the vehicle will be purchased; *Budget*, which outlines the financial range for the purchase; and *Brand*, which indicates the preferred manufacturer. Each of these classes is designed to precisely align the vehicle offerings with the user's individual expectations and requirements.

It's crucial to acknowledge that in different domains and applications, additional preferences may be required to comprehensively address user needs. Consequently, subclasses of the *Preference* class can be tailored and customized to align with the specific demands of the vehicle sales domain and its user base. By integrating these preferences into the user contextual profile ontology, we can provide a more personalized and tailored user experience.

Using the conceptual model, we delineate the principal classes, attributes, and relationships among them (as depicted in Figure 2 and Figure 3). The implementation phase of the UCPO involves the actual creation and deployment of the ontology within a system or application. This necessitates transforming the ontology's conceptual model into a computer-readable format utilizing ontology languages like OWL or RDF. Figure 4 demonstrates the implementation of the UCPO using OWL, with support from the Protégé-OWL editor (Musen, 2015). The OWL version of the UCPO ontology is publicly available online[2], allowing researchers to inspect, reuse, or extend the ontology for further development and integration.

To operationalize LLMs within the ontology engineering process, we adopt a structured pipeline comprising three stages: (i) input preparation and prompt design, (ii) LLM querying, and (iii) post-processing and integration. The prompt engineering component is essential for shaping LLM behaviour. We employ a range of prompting strategies - including zero-shot, few-shot, and chain-of-thought prompting. For instance, chain-of-thought prompting helps decompose complex competency questions into semantic components, while few-shot prompting guides the generation of property hierarchies using examples from existing ontologies. These strategies are implemented using *OpenAI's GPT-4o API*, accessed via Python scripts with structured prompt templates. The generated outputs are parsed and systematically mapped to OWL axioms using the *RDFLib* and *Owlready2* Python libraries.

---

[2] https://github.com/lengocluyen/ucpo_okb

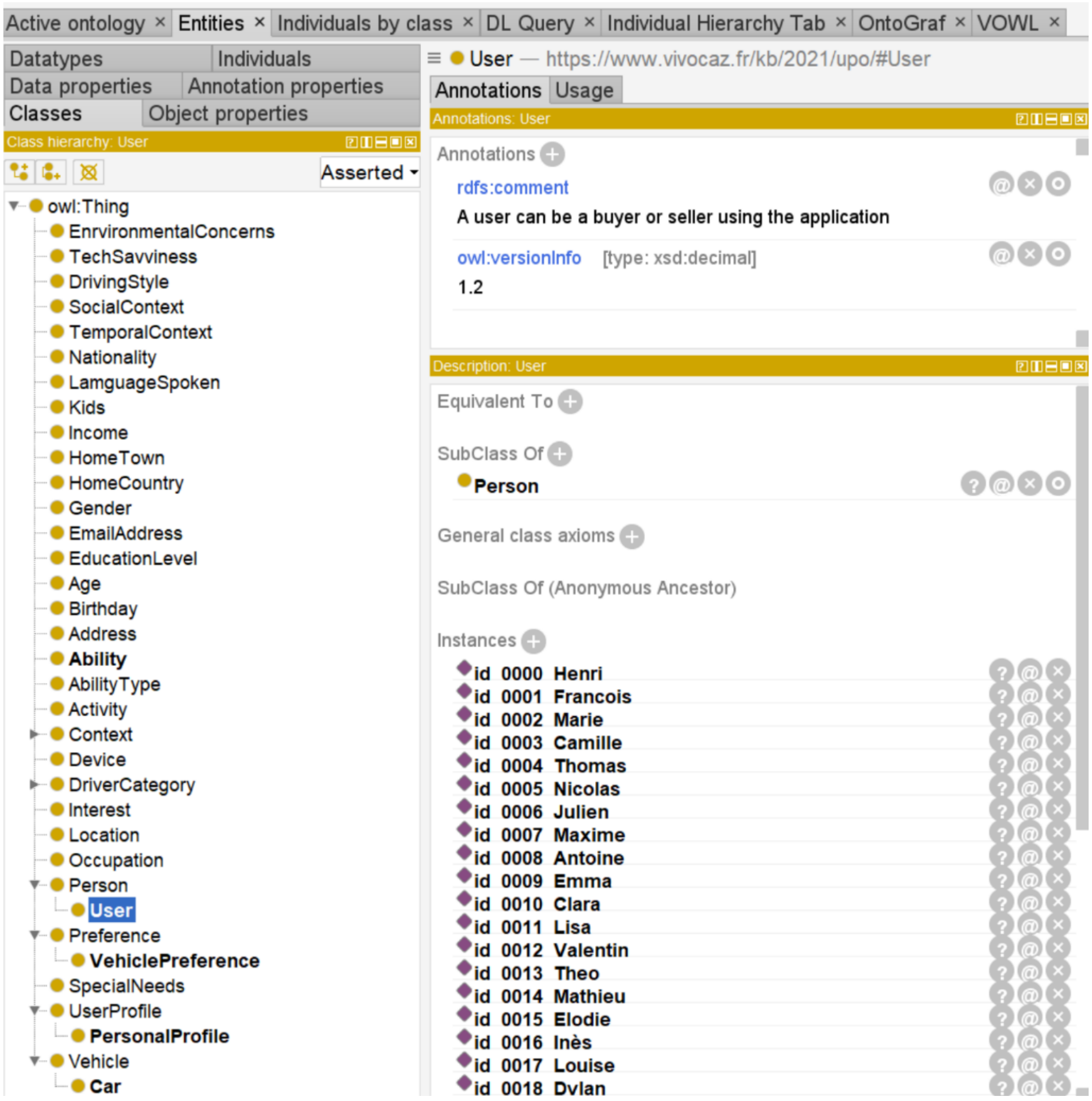

Figure 4: A snapshot of the User Contextual Profile Ontology in the Protégé-OWL editor

After developing the modelets, the next step involves subjecting them to a test case generation process to determine their effectiveness and accuracy before their integration into the primary ontology model. During this phase, a series of test cases is formulated based on predefined competency questions and requirements. These test cases are designed to assess how well the modelets function and perform in addressing specific scenarios and user needs. Through rigorous testing, any potential gaps or inconsistencies in the modelets can be identified and resolved before their integration into the main ontology model. Subsequently, document generation achieves a reduction in time by using LLMs to generate comments and annotations about classes and relationships. The final step in our proposed methodology is a feedback loop that collects comments from stakeholders or end users. The following section will focus on the experiments and evaluations of the obtained ontology, assessing its effectiveness and structure in detail.

### *4.3 Evaluation Metrics and Results*

To rigorously evaluate our method, we adopt a combination of model-level, data-level, and query-level assessments. We justify this multi-level evaluation framework as it not only verifies logical consistency and structural design, but also tests the ontology's applicability with realistic data instances and SPARQL-based information retrieval. These evaluation dimensions are essential to ensure both theoretical soundness and practical relevance in dynamic applications like vehicle recommender systems.

Table 4: Results achieved from applying various metrics to UCPO

| **Base Metrics** | |
|---|---|
| Class count | 42 |
| Object property count | 31 |
| Data property count | 16 |
| Properties count | 47 |
| Individual count | 159 |
| *SubClassOf* axioms count | 11 |
| Object property domain axioms count | 30 |
| Object property range axioms count | 30 |
| DL expressivity | $ALH(D)$ |
| **Schema Metrics** | |
| Attribute richness ($AR$) | 0.380952 |
| Inheritance richness ($IR$) | 0.261905 |
| Relationship richness ($RR$) | 0.738095 |
| Axiom/class ratio | 27.547619 |
| Class/relation ration | 1.0 |

We experiment with testing and evaluating the produced ontology. Generally, the tests can be categorized into model tests, data tests, and query tests. To perform the model test, we employ evaluation metrics such as OOPS! (Poveda-Villalón, et al., 2014), OntoQA (Tartir, et al., 2010), and OntoMetrics (Lantow, 2016) to verify the overall consistency of the developed ontology. The data test involves checking the validity of the model after populating it with instance triplets. Finally, for query tests, ICQs must be transformed into SPARQL queries to ensure that the expected answers are obtained. The ontology model must be adjusted until all tests are successful. This section presents our method with the support of LLMs, specifically in populating instance triplets and generating SPARQL queries based on ICQs to conduct these types of tests and ensure the ontology's adaptation.

In our study, we assessed the quality of our ontology by analysing its structure using various metrics from the OntoMetrics framework. The results, detailed in Table 4, include both base and schema metrics. Base metrics, which are straightforward and quantitative, tally the number of axioms, classes, object properties, data properties, and individual instances within the ontology. These metrics offer insights into the ontology's compositional elements. The analysis suggests that our ontology is a lightweight knowledge graph suitable for various application architectures. Furthermore, the ontology's Description Logic (DL) expressivity is classified as $ALH(D)$, indicating that it utilizes attribute language $AL$ with a role hierarchy ($H$). This classification provides a deeper understanding of the ontology's logical framework.

In the realm of schema metrics, the evaluation of the ontology centres on assessing attribute, inheritance, and relationship richness (Tartir, et al., 2010). Specifically, attribute richness ($AR$) is defined as the average number of attributes per class, calculated as follows:

$$AR = \frac{|NA|}{|C|} \tag{1}$$

where $NA$ represents the total number of attributes across all classes, and $C$ signifies the number of classes. A high attribute richness score reflects a superior quality of ontology design and indicates a significant amount of information associated with instances. Meanwhile, inheritance richness ($IR$) is quantified as the average number of subclasses per class, calculated as follows:

$$IR = \frac{|H|}{|C|} \tag{2}$$

where $H$ represents the total number of inheritance relationships within the ontology. Inheritance richness provides a measure of how information is distributed across various levels of the ontology. Additionally, relationship richness ($RR$) is defined as the percentage of relationships that exist between classes, and it is calculated as follows:

$$RR = \frac{|P|}{|H| + |P|} \tag{3}$$

where $P$ denotes the total number of non-inheritance relationships within the ontology. Inheritance richness serves as a metric to evaluate the variety of relationship types present in the ontology.

Additionally, other metrics like the axiom/class ratio and class/relation ratio illustrate the relationships between axioms and classes, and between classes and relations, respectively. The results of these metrics, detailed in Table 4, suggest that our proposed ontology strikes a balance between being a horizontal (or shallow) ontology and a vertical (or deep) ontology.

To perform a data test on the UCPO, it is necessary to verify the model's validity after it has been populated with instance triples. Consider, for instance, that we have developed two profiles for *Henri*: a professional profile and a family profile. We could fill his professional profile with data like his job title and workplace, and his family profile with details such as the number of children he has and his favourite family activities. This process confirms that the UCPO accurately and coherently represents this information.

To carry out a query test on the UCPO, we need to transform the ICQs listed in Table 3 into SPARQL queries. This ensures that we retrieve the anticipated answers. The queries must be designed to efficiently extract pertinent information from the ontology, making use of the relevant classes, properties, and instances. For example, if the competency question is *"What is the user's preferred vehicle brand?"*, an appropriate SPARQL query can be constructed as shown in Query Box 2.

```
PREFIX vo: <http://vivocaz.fr/vo/ns#>
PREFIX ucpo: <http://vivocaz.fr/ucpo/ns#>
SELECT ?user ?brand
WHERE {
 ?user ucpo:hasUserProfile ?userProfile.
 ?userProfile ucpo:hasVehiclePreference ?userVehiclePreference .
 ?userVehiclePreference upo:hasFavoriteBrand ?brand .
} ORDER BY ?user LIMIT 10
```

Query Box 2: The SPARQL query expression used to search for the first 10 users and their favourite brands

```
PREFIX ucpo: <http://vivocaz.fr/ucpo/ns#>
SELECT ?vehicleModel ?efficiency
WHERE {
 ?profile ucpo:hasDrivingPurpose "professional" .
 ?profile ucpo:hasVehiclePreference ?vp .
 ?vp ucpo:hasFuelEfficiency ?efficiency ;
 ucpo:recommendsVehicle ?vehicleModel .
  FILTER(?efficiency > 30)
}
```

Query Box 3: The SPARQL query to retrieve fuel-efficient vehicles suitable for professional users

To further demonstrate the utility of the UCPO in the vehicle sales domain, the ontology was populated with a range of individual instances reflecting diverse user profiles, preferences, and contextual factors. For example, one profile represents a user named *Henri*, who maintains two contextual profiles: one tailored for professional commuting and another for family travel. *Henri's* professional profile includes preferences such as fuel efficiency, compact size, and compatibility with electric vehicles. In contrast, his family profile emphasizes safety features, larger seating capacity, and enhanced comfort. These variations allow the ontology to model nuanced user needs and retrieve context-appropriate vehicle recommendations. To assess retrieval accuracy, we issued SPARQL queries aligned with competency questions as shown in Query Box 3.

Table 5: Summary of Leveraging LLMs in the development of the UCPO

| **Stage** | **Description of LLM Usage** | **Impact** |
|---|---|---|
| Scenario and Glossary | Analyse research papers to identify key concepts related to user context and personalization in the automotive industry. | Informs the creation of a glossary of terms relevant to the user profile and contextual data. |
| Informal Competency Questions | Assist in generating a list of ICQs by analysing user interaction patterns and preferences within the automotive domain. | Expands the scope of questions considered, leading to a more robust UCPO. |
| Modelet Development | Suggest new concepts, relations for example *DrivingStyle*, *TechSavviness*, or *Environmental Concerns*. | Enhances the comprehensiveness of the user profile by capturing nuanced user characteristics. |
| Test Case Generation | Assist in transforming ICQs into SPARQL queries for testing purposes. | Expedites the testing process and improves efficiency. |
| Model Refinement | Analyse the outcomes of various query forms to identify areas for improvement in the ontology, including missing attributes or relationships. | Optimizes the ontology's information retrieval capabilities and ensures its effectiveness in real-world applications. |
| Document Generation | Generate comments and annotations about classes and relationships within the ontology. | Reduces time required for ontology documentation. |
| Feedback | Process large volumes of textual feedback from stakeholders. Extract key themes, sentiments, and suggestions. Identify recurring patterns or issues. | Prioritizes improvement efforts based on feedback. Identifies areas of consensus or disagreement. Generates recommendations for ontology refinement. |

The query returned five distinct vehicle models (e.g., Renault Zoe, Peugeot e-208) aligned with *Henri's* efficiency threshold. Additional queries focused on cross-context matching - identifying vehicles that satisfy both professional and family requirements - revealed only one model satisfying both contexts (Peugeot 5008 Hybrid), validating the ontology's capacity to support multi-profile personalization.

By conducting model, data, and query tests, we thoroughly assess an ontology's structure, the accuracy of its populated instances, and its capability to respond to competency questions, ensuring both theoretical adherence and practical efficacy. Model tests scrutinize the ontology's logical coherence, checking for alignment with the domain model and resolving any inconsistencies. Data tests examine how well the ontology handles real-world data, crucial for its utility in practical scenarios. Query tests, enhanced by LLMs, automate the translation of ICQs into precise query languages like SPARQL, testing the ontology's information retrieval capabilities. LLMs also generate diverse query forms and analyse outcomes, speeding up refinements necessary for the ontology's functionality. This integration of LLMs into the testing process not only improves efficiency but also ensures the ontology's reliability and relevance in dynamic application environments, making it a robust tool for developers. In the next section, we discuss some of our findings regarding the advantages and drawbacks of leveraging LLMS for the development of ontological knowledge bases.

Finally, the use of LLMs in developing the UCPO enhances the efficiency and effectiveness of the ontology engineering process. By automating key tasks, facilitating deep insights from data analysis, and enabling adaptive and personalized applications, LLMs have proven to be powerful tools in the development of a robust and flexible ontology tailored to user context profiles. Assistance at each stage is summarized in Table 5. In general, the case study in our work highlights several advantages of using LLMs in ontology development. Firstly, LLMs can efficiently analyse extensive literature corpora, extracting relevant concepts, terms, and information, thereby significantly reducing the manual effort involved in literature reviews. They also suggest new classes, properties, or relations based on their understanding of the domain, which can enhance the ontology's coverage and completeness. Additionally, LLMs facilitate the documentation process by automatically generating comments and annotations for classes and relationships, thereby reducing the time and effort required. They also assist in generating test cases, including populating the ontology with instance triplets for data testing and converting ICQs into formal queries for query testing. Furthermore, LLMs support the iterative refinement of the ontology by generating diverse query forms and analysing their outcomes, ensuring the ontology can accurately retrieve relevant information.

## 5. Discussion

The integration of LLMs into the ontology engineering lifecycle, as demonstrated through the development of the UCPO, has unlocked new avenues for accelerating and enhancing knowledge acquisition and representation processes. We highlight the advantages of leveraging LLMs in ontology development. With their remarkable natural language understanding capabilities, LLMs enable us to efficiently extract relevant concepts, entities, and relationships from diverse data sources, including unstructured text, thus accelerating ontology population and enrichment. Moreover, by utilizing LLMs' language generation abilities, we can automate the creation of glossaries, ensuring consistent terminology usage and reducing ambiguity across the knowledge base. Additionally, LLMs facilitate the generation of documentation, test cases, and other artifacts, enhancing the ontology's interpretability, maintainability, and adoption. Furthermore, the language understanding and generation capabilities of LLMs facilitate iterative refinement cycles, allowing us to efficiently identify gaps, inconsistencies, and opportunities for ontology enhancement.

The adoption of LLMs for ontology development introduces a range of complex challenges that require careful consideration by domain experts and ontology engineers. Firstly, related to issues of ontological consistency and coherence, we find that LLMs are proficient in pattern recognition and language generation. However, their outputs often contain hallucinations and lack the logical consistency and coherence required by formal ontological structures, making human oversight and post-processing essential to maintain the integrity of the knowledge base. Moreover, the integration of domain-specific expertise remains crucial. LLMs rely on training data collected from the internet that may not fully capture the nuances and complexities of specialized fields. Therefore, close collaboration with subject matter experts is necessary to ensure the validity and refinement of LLM outputs. Additionally, scalability and performance become significant as ontologies expand in size and complexity. The scalability of LLM-based methods may become a constraint, prompting the need for optimization strategies or the integration of hybrid methodologies that combine LLMs with traditional knowledge representation techniques. Lastly, transparency and explainability remain essential. LLMs often function as black boxes, making it difficult to understand and articulate their decision-making processes. Enhancing the interpretability and transparency of these models is crucial for building trust and ensuring accountability in LLM-assisted ontology development.

Moving forward, a balanced and pragmatic approach is recommended, where LLMs are leveraged judiciously to augment and accelerate ontology engineering processes, while ensuring close collaboration with domain experts and adherence to best practices in knowledge representation and reasoning. Additionally, ongoing research efforts should focus on addressing the challenges of ontological consistency, debiasing, scalability, and model transparency, ultimately paving the way for more trustworthy and reliable LLM-driven ontology development pipelines.

## 6. Conclusion & Future Work

In this paper, we delve into the development of OKBs by leveraging the capabilities of LLMs. Our proposed methodology seamlessly integrates LLMs into ontology engineering workflows, where they play a dual role as stakeholders and tools. The ontology development process outlined in our methodology comprises seven essential steps, including scenario and glossary creation, ICQs formulation, modelet development, test case generation, model refinement, documentation generation, and feedback incorporation. At each stage, LLMs demonstrate their versatility by leveraging their natural language capabilities to facilitate knowledge acquisition, automate artifact generation, and facilitate iterative refinement cycles. To validate and experiment our proposed methodology, we conducted a case study focusing on the development of a user context profile ontology. This case study illustrates how each step of the methodology contributes to the creation of an OKB, as introduced in the scenario. By leveraging LLMs' ability to extract relevant concepts and relationships from heterogeneous data sources while automating documentation and test case generation, our approach significantly expedites ontology population and enhancement. However, it is imperative to address challenges such as ensuring ontological consistency, achieving domain coverage, mitigating bias, and tackling scalability and interpretability issues.

We acknowledge the inherent limitations of LLMs, such as their susceptibility to biases and challenges in maintaining factual consistency or domain accuracy. Thus, we advocate for a hybrid approach that leverages LLM capabilities alongside expert-guided validation and formal knowledge engineering principles. Future work should prioritize improving the transparency and interpretability of LLM-generated ontological content, incorporating debiasing mechanisms, and scaling methods to accommodate large, evolving domains with minimal manual intervention. Beyond technical refinement, further research can explore domain-specific applications of LLM-assisted ontology development in areas such as personalized learning systems, semantic search engines, intelligent digital assistants, and medical knowledge bases. Another promising direction involves the integration of LLMs into continuous ontology evolution workflows, where models assist in updating ontologies in response to real-time data streams or evolving domain needs. Additionally, benchmarking and evaluating the quality of LLM-generated ontologies across diverse domains remain open challenges, requiring standardized metrics and robust evaluation protocols. Ultimately, future efforts should not only optimize methodologies but also broaden the scope of deployment, enabling the dynamic co-construction of ontological knowledge in interdisciplinary and multilingual settings.